# Deep-Learning-based Vulnerability Detection in Binary Executables

Andreas Schaad[1] and Dominik Binder[1]

Offenburg University of Applied Sciences, Germany
{andreas.schaad,dominik.binder}@hs-offenburg.de

**Abstract.** The identification of vulnerabilities is an important element in the software development life cycle to ensure the security of software. While vulnerability identification based on the source code is a well studied field, the identification of vulnerabilities on basis of a binary executable without the corresponding source code is more challenging. Recent research [1] has shown, how such detection can be achieved by deep learning methods. However, that particular approach is limited to the identification of only 4 types of vulnerabilities. Subsequently, we analyze to what extent we could cover the identification of a larger variety of vulnerabilities. Therefore, a supervised deep learning approach using recurrent neural networks for the application of vulnerability detection based on binary executables is used. The underlying basis is a dataset with 50,651 samples of vulnerable code in the form of a standardised LLVM Intermediate Representation. The vectorised features of a Word2Vec model are used to train different variations of three basic architectures of recurrent neural networks (GRU, LSTM, SRNN). A binary classification was established for detecting the presence of an arbitrary vulnerability, and a multi-class model was trained for the identification of the exact vulnerability, which achieved an out-of-sample accuracy of 88% and 77%, respectively. Differences in the detection of different vulnerabilities were also observed, with non-vulnerable samples being detected with a particularly high precision of over 98%. Thus, the methodology presented allows an accurate detection of 23 (compared to 4 [1]) vulnerabilities.

## 1 Introduction

### 1.1 Background

Identifying vulnerabilities is an important element of the software development process to ensure the security of software. In the early stages of development, this can be done by testing the code and performing static analysis based on the source code. The identification of vulnerabilities, however, becomes more challenging when analysing applications without knowledge of the associated source code. This usually occurs when analysing legacy applications, proprietary software or other forms of black-box pentesting scenarios. In these cases, black-box tests can be used to identify vulnerabilities based on the behaviour of an application without knowing its internal workings.

Unfortunately, black-box analysis methods have a number of disadvantages. Vulnerability detection methods such as fuzzing are very time-consuming, offer low code coverage and have high resource requirements. Furthermore, the success of these methods depends heavily on the specification of the test and the completeness of the test cannot be proven [2]. Since in an analysis scenario without the presence of the source code, the application would still be available in the form of assembly code, this code could be used for an analysis to avoid the disadvantages of a black-box analysis. However, due to its complexity a manual analysis of assembly code performed by humans may hardly be feasible for larger applications.

The analysis of assembly code is therefore particularly interesting in the form of an automated analysis. Since the creation of a program for automated analysis is not a trivial task due to the complexity of the code, this demanding programming task in the field of binary code analysis can be accomplished by using machine learning techniques [3], whereby statistical models are used to identify relevant associations for the detection of vulnerabilities on the basis of a database of already known vulnerable samples.

### 1.2   Motivation

The use of machine learning for processing program code opens up a variety of interesting application areas. Besides promising application areas such as automated code generation [4], where impressive progress has been made recently, the analysis of existing code is a widely studied area [3].

An analysis of existing source code on the basis of machine learning can be used as a supplementary or alternative means of a conventional static code analysis to find bugs within the code. The advantage of this approach is that no fixed set of rules has to be defined for this static analysis, but it can be learned from a database by the model used. The application of machine learning based analysis methods to binary files is more challenging. A term used in this context is binary code similarity analysis, in which characteristics of different samples are compared [5]. Using this method as a foundation already supports a variety of applications. In particular, the reuse of code snippets can be detected. Such a detection is of interest when determining authors and the reuse of identical functions of programs [6], a technique used in particular in malware analysis [7]. Similarity-based analysis methods can also be used for the discovery of vulnerabilities, which is useful if an already known vulnerability is reproduced in a nearly identical form, for example by reusing a complete function [8].

In order to offer a generalised form of vulnerability detection for binaries, the use of more profound forms of analysis is necessary. Such an analysis would not only take into account the similarity with already known patterns, but also include the actual inner workings of the code under investigation. Research into machine learning-based fuzzing methods has already shown that they are suitable for significantly improving the runtime and success rate of the vulnerability discovery process [9], [10]. While such an approach is more effective, it still remains dependent on the dynamic execution of the program and, despite a good coverage of the program flow, cannot guarantee completeness and claims a certain runtime, which however is shorter compared to conventional fuzzing.

In order to avoid execution of the code and the associated problems mentioned, the approach of a machine learning based static analysis of binary code is an appealing option. For this purpose, we define the analysis of binary code as any form of analysis of information that we can extract from a binary executable. A decompilation into LLVM Intermediate Representation would be an example of this. In [1], this approach was demonstrated using deep learning in the form of recurrent neural networks. The authors used a selection of six architectures to perform a classification of four vulnerabilities, which were categorised by using the Common Weakness Enumeration (CWE). Also, instead of directly using assembly language as input for the models, the code was decompiled into LLVM Intermediate Representation, which we equally use as our working definition of binary code.

### 1.3 Aims and Objectives

This paper aims to analyse whether deep learning-based models can be used to sufficiently identify vulnerabilities (categorised by CWEs) in binary executables. The adopted methodology is based on the approach from [1], and extends this in both implementation and scope. Therefore, the following objectives are examined:

- Can the approach shown in [1] be reproduced and significantly extended beyond identification of only four vulnerabilities?
- To which of the 118 vulnerability types (categorised by CWEs) in the used dataset (SARD) can the approach be extended and how well are they identifiable?
- Is such a model able to identify the exact type of vulnerability or is it more efficient at identifying the presence of an arbitrary vulnerability?
- Which architectural design decisions influence the result?

Based on the defined objectives, this paper first provides an overview of related research (Section 2). Then, the creation of the dataset used and the training process based on that data is described in detail (Section 3 and Section 4). Afterwards we discuss our findings, its implications and possible improvements (Section 5 and 6).

## 2  Related Work

Previous work on machine learning based vulnerability detection can be divided into static versus dynamic analysis approaches. When considering static analysis methods, those can be further divided on static analysis of the source code and analysis techniques in which no source code is available.

### 2.1   Static vulnerability detection based on source code

In the area of vulnerability detection through static source code analysis, research has been conducted to determine a code similarity order of a fingerprint to already known vulnerabilities. A number of publications have shown how algorithmic solutions can be used to detect reused vulnerable code fragments on the basis of the source code [11], [12], [13], [14]. Limitations of these approaches were found in the detection of new vulnerabilities that were not a direct copy of known vulnerabilities or were heavily modified.

A detection of new vulnerabilities requires a deeper understanding of the code to be analysed, which can be implemented by deep learning based analysis. VulnDeePecker [15] uses such a deep learning based detection method on program slices to detect bidirectional LSTM neural network API function calls related vulnerabilities. For this purpose, six datasets were derived from the SARD dataset [16] and tested with a set of deep learning architectures of different depths. It was determined that architectures with two to three layers performed best, resulting in an F1-score of 86.6 - 95 % depending on the dataset.

In VulDeeLocator [17] it was demonstrated how the use of an intermediate code representation and an associated reduction of the code of interest can be used for detection. It was shown how the use of the intermediate code representation enabled the detection of vulnerabilities with an F1-score of 90.2 % to 96.9 %.

Further work in the area of source code based detection showed how vulnerabilities on a function-level can be detected by using abstract syntax trees [18], [19].

### 2.2   Static vulnerability detection based on binary executables

Another subarea of static analysis is in the area of vulnerability detection without using the source code, which will also be the focus of this work. Instead of processing the source code, this work involves extracting relevant information from the binary executable.

In this scenario, the detection based on code similarity was also investigated. In [20], it was shown how similarity-based binary detection methods can be used under consideration of cross-architecture. For this purpose, binary code from ARM, MIPS and x86 CPU architectures was analysed by first translating them into an intermediate representation and later deriving a similarity score of the translated samples.

Another similarity-based detection method that does not require the use of source code is discovRE [21]. This tool uses a k-Nearest Neighbour algorithm to identify similar functions based on numerical features and then filter them based on similarity of control-flow graphs. A special feature of this work is the efficiency of the tool, which can check over 130,000 functions in 80 milliseconds. However, the control-flow graph-based filtering of relevant functions is also one of the major drawbacks of this approach, since minor changes in the control-flow makes detection very difficult.

Beyond the similarity-based detection approach, the use of deep learning methods is also an option for the processing of binary executables. The authors of [22] showed how a detection based on decompiled ASM code can be performed. Similar to the deep learning methods based on the source code, the transformed code was processed by recurrent neural networks following methods of natural language processing (NLP). In their work, only stack-based buffer overflows were considered, which were collected from public repositories, which is why these are very realistic data. The authors report that their model achieved perfect classification results both in-sample and out-of-sample, although it cannot be ruled out that this is due to intensive hyperparameter optimization.

In addition to the use of assembly code, it is also possible to use an intermediate representation of the code for this case, as demonstrated in [1]. In this work a total of 14,657 code fragments were used for the four vulnerability types CWE-134 (Use of Externally-Controlled Format String), CWE-191 (Integer Underflow (Wrap or Wraparound)), CWE-401 (Improper Release of Memory Before Removing Last Reference ('Memory Leak')) and CWE-590 (Free of Memory not on the Heap) of the SARD dataset. These code fragments consist of LLVM Intermediate Representation code extracted from the compiled binary executables.

To classify the different vulnerabilities, six different variants of recurrent neural networks were trained with the bidirectional simple recurrent neural network giving the best result. The results were only presented graphically but based on the reported accuracy, a range of 98 - 100% could be deduced. Since this work serves as an orientation to this paper, the implications of this work will be discussed in more detail later in our work.

### 2.3 Dynamic vulnerability detection based on binary executables

In addition to the use of machine learning for the static detection of vulnerabilities, the increasing use for the optimization of dynamic methods is also worth mentioning. For this, research showed how existing fuzzing methods can be improved by machine learning in terms of detection rate and code coverage.

In [23] it was shown how machine learning can be used to optimize the seed inputs for the fuzzing process carried out on four PDF viewers, in order to reach more execution paths. The improved seed generation of the fuzzer resulted in 32% more execution paths.

NeuFuzz [24] follows a similar approach and, in addition to the improved seed selection strategy, also shows how deep learning techniques can be used to predict whether an execution path is vulnerable. Using real-world programs, the authors have shown that this approach can detect more vulnerabilities in less time than conventional fuzzing tools.

## 3 Data gathering and preprocessing

This section contains the preparation of a dataset for a training process to create a deep learning-based detection of vulnerabilities in binary executables. Since this process requires excessive preprocessing and preparation of the data, the whole machine learning process carried out in this paper is divided into the description of the dataset in this section separately from the actual training process following in Section 4.

### 3.1 Overview

The approach adopted in this paper follows the methodology reported in [1], where we initially replicate the reported results and then extend their approach to cover 19 additional CWEs and required preprocessing steps. This approach can be defined using the taxonomy introduced in [3] as supervised deep learning using recurrent neural networks for the application of vulnerability discovery. Therefore, code-base features on a token-level are embedded by using a word-to-vector model.

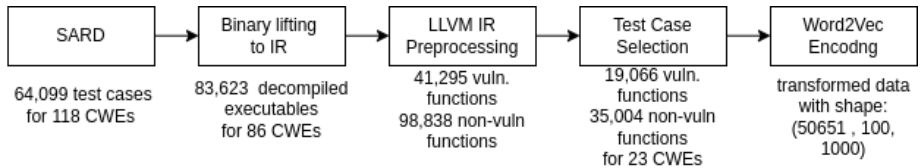

**Fig. 1.** Data gathering and preprocessing steps

The steps for generating and preprocessing our dataset follow those of [1] and are outlined in Figure 1 and detailed in the following.

### 3.2 Software Assurance Reference Dataset

The Software Assurance Reference Dataset[1] published by the National Institute of Standards and Technology contains code of over 170,000 programs in the programming languages C, C++, Java, PHP and C#. The programs included represent test cases taken from a selection of 150 weaknesses, following the Common Weakness Enumeration (CWE). The test cases are labelled "good" or "bad", indicating within the dataset whether the respective weakness is present [16].

This means that both non-flawed and a flawed implementation of the respective test cases are included, which is why the use of this dataset is considered to be suitable for training a vulnerability detection model. As part of the SARD subset, the Juliet Suite for v1.3 C/C++ contains 64,099 artificially generated test cases[2] that will be used in this work.

---
[1] https://samate.nist.gov/SARD/

The scope of the included programs primarily covers the use of the standard C library API for all platforms. Beyond that, no other third-party libraries are used in the dataset [25].

The dataset contains weaknesses in the form of CWEs. However, since our work focuses on vulnerability detection, we want to define those terms in more detail. When speaking of vulnerability detection in this context, a vulnerability is represented as the result of one or more weaknesses [26]. Conversely, despite the presence of a weakness, filtering an input may not result in an exploitable vulnerability [16]. Since we are using a function-based approach, we only consider the filtering of input within a function. We therefore speak of vulnerability detection in this context, even though this dataset only contains weaknesses. In other words, we examine functions in isolation to see if they have a vulnerability, even if the vulnerability found is potentially non-exploitable by filtering outside that function.

### 3.3 Dataset Preparation

The code of the dataset is available in the form of source code divided into different CWEs and test cases. However, since we do not want to process the source code for further processing, but rather in the form of a black-box-based approach, the code first gets compiled into binary executables. This is performed automatically using already public code from the [27] repository, which takes care of the organisation of the code, generation of the correct make files, separation of the compiled files by CWE type and the presence or absence of a weakness.

### 3.4 Decompilation

With the compiled executables created as a starting point, the analysis scenario corresponds to a real scenario without available source code. A decompilation to LLVM IR using the tool *RetDec*[3] is performed using its default configuration, as it was also used in [1]. We expect from this preprocessing step that useful properties of LLVM IR will simplify the learning of relevant structures of the code in the later training steps.

Since all Windows-specific CWEs were excluded in the previous step, these are test cases from 86 remaining CWEs. The compilation process resulted in the separate storage of flawed and non-flawed implementations of individual test cases, which is why the total number of files to be processed increased to 83,623 executable files. We further noticed unintended over-optimisation effects performed by RetDec, resulting in simplified test cases, which no longer contian the flawed implementation. With regard to these cases, we were able to ensure that they were completely removed through the subsequent selection processes.

---

[2] https://samate.nist.gov/SARD/resources/releaseJuliet1.3Doc.txt
[3] https://github.com/avast/retdec

### 3.5  LLVM IR Preprocessing

The following preprocessing serves to put the textual information of the decompiled functions into a standardised form that can be used for further processing Therefore, a custom parser was implemented, which also performs the following standardisation steps on each file:

- Global variables, local variables and labels are standardised by replacing each unique keyword with a prefix and a counter variable, so that these keywords are then present in the form of "LBL_1", "LBL_2", "VAR_1", "VAR_2", etc.
- Function calls are distinguished between local and non-local functions. Non-local functions come from external libraries and already have a unique name across several applications. For example, a standard C function like "memcopy" has the same name across all applications, which was generated by RetDec. We deliberately leave these in the code and do not standardise them, as the recognition of these functions commonly used for programming is relevant for finding vulnerabilities. Local functions are instead replaced with a fixed keyword and are therefore identically for all local functions, since those do not contribute to the logic of a vulnerability.
- All numbers get standardised by splitting each numeric value into individual digits, which are treated as individual tokens.
- An "EOL" (end of line) token is added after each code line, to preserve the contextual information of individual code lines even after combining them into one large array.
- The processed code lines are split to individual tokens which are represented by one array of tokes per function processed.

This process was carried out for all 83,623 files and 86 CWEs in total. As a result of this step those functions are now available in form of standardised arrays of tokens, which are used for the following steps. An overview of the number of all extracted test cases is given in Appendix A.

### 3.6  Test Case Selection

The number of the extracted functions is narrowed down to a number of relevant weaknesses, based on the following criteria:

- The windows specific weaknesses are excluded
- From the selection of all functions of the test cases, only those are selected which are directly related to the respective weakness
- All weaknesses with less than 500 test cases are excluded
- All samples with a length below 300 tokens are removed
- Only weaknesses which are also present after compilation are considered (e.g. CWEs like "CWE 615, Information Exposure Through Comments" are removed)

For this purpose, using the terminology defined for the SARD dataset, only the relevant primary and secondary good functions as well as the primary good functions were selected by using regular expressions. The final selection of 54,070 samples from 23 CWEs is listed in Appendix B sorted by descending absolute frequency. The percentage proportion of individual weaknesses ranges from 0.61 % to 12.09 %. Due to the chosen order of selection, there is also one CWE with less than 500 entries, which is, however, negligible for this individual class.

### 3.7 Word to Vector Encoding

To transform the data, in the form of arrays of tokens, into a vector, a Word2Vec model using the continuous bag-of-words architecture is trained on the created dataset using the python *gensim* library. For this purpose, all tokens of the 54,070 samples are combined to a corpus of 30,710,959 tokens in total, containing 760 unique tokens. The model was trained with parameters of 100 dimensions, a context size of 3 and a downsampling rate of 1e-3. Subsequently, all samples were transformed token by token using the trained Word2Vec model and a zero-padding to a sequence length of 1000 was carried out. In [28] it was shown that with LSTM architectures, pre-padding provides significantly better results, which is why pre-padding is used in this work.

## 4  Machine learning - Training and Evaluation

Two experiments of vulnerability detection are performed for a selection of different machine learning architectures (Section 4.2). The first experiment is a binary classification, which is only intended to detect whether any vulnerability is present. For the second experiment, a classification is performed on the specific vulnerability, which should determine the exact nature of the vulnerability.

### 4.1 Dataset and Training Architecture

We apply a train-test-validation split to the total of 50,651 samples, with 70/15/15 % of the data respectively. The test data was used in the following training process for callbacks for training optimisation, which is why we use the additional validation split for a final out-of-sample evaluation to exclude overfitting.

The used architectures are partially based on [1] and consist of recurrent neural network layers followed by ordinary fully connected feed forward layers. However, as we specifically evaluated the learning effects for larger neural networks, LSTM, GRU and SRNN architectures were also applied. Each architecture was tested as a unidirectional and bidirectional variant. For each variant, a smaller single layer and a two layer variant were tested to check how the model reacts to a larger capacity. This size refers to the number of RNN layers, as well as the subsequent fully connected layers. A unit size of 64 is used for all RNN layers, as well as the feed forward layers. All RNN layers use a *tanh* activation function, all fully connected layers use a linear activation function.

### 4.2   Training and Model Selection

This section details the training process for the binary and multi-class classification, which are described below. Thereby, we aim to find out which architecture is most suitable in solving the given problem and which design decisions take influence on the result.

All training tests are carried out using a batch size of 64, to optimise the categorical cross entropy using an Adam optimiser with an initial learn rate of 1e-4. The learning rate was automatically decreased by half using the *ReduceLROnPlateau*-callback if the validation loss did not improve for five epochs in a row. A training run was interrupted by an early stopping callback if the validation loss did not improve after 15 epochs. The classes were weighted for the training process according to their frequency to compensate for the imbalance.

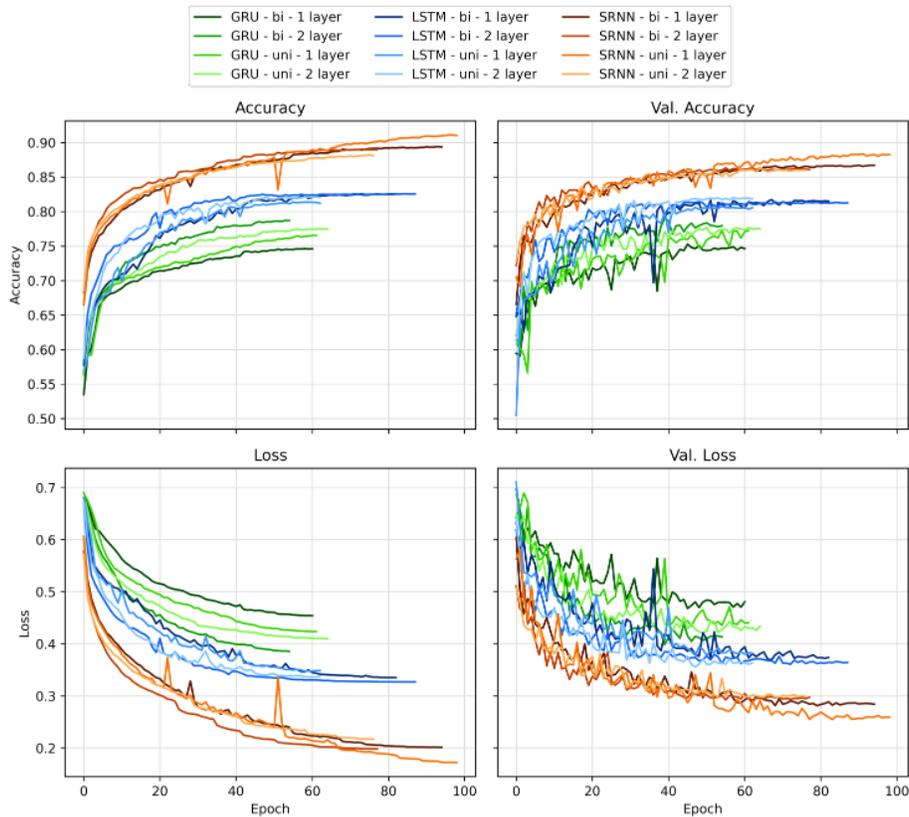

**Fig. 2.** Accuracy and loss for the binary classification training process

**Binary Vulnerability Classification** The implementation of binary classification describes the experiment in which only a classification between a non-flawed (class 0) and a flawed (class 1) implementation is carried out, without considering the exact nature of the vulnerability.

Figure 2 shows the results of the tested architectures across all training epochs. Based on the data, we see a successful training process across all architectures, which is also evident out-of-sample without overfitting. The group of SRNN models produced the best results in this respect, with the differences between the various variants within the group being low. Out of this group the single layer unidirectional SRNN model achieves the best accuracy and lowest loss both in-sample and out-of-sample.

Since the use of callback functions might influence the shown validation results, a separate true out of sample using untouched validation data (as described in Section 4.1) was used to further validate the selected single layer unidirectional SRNN model. A detailed interpretation of the test results is given in Table 1. In addition to the overall accuracy of around 88 %, this table also shows the imbalance of the classes, the F1-scores per class, as well as the precision and recall metrics. From the frequency of the classes (the support column of the table), a baseline of 65.86 % accuracy is obtained. This baseline was largely exceeded by the selected model, as well as by all other models. With a precision of 96% for the detection of non-flawed implementations, it can be stated that the model is very well suited to identify flawless implementations, despite the existing false positives.

| Class ID | Precision | Recall | F1-score | Support |
|---|---|---|---|---|
| 0 | 0.96 | 0.85 | 0.91 | 5004 |
| 1 | 0.77 | 0.94 | 0.84 | 2594 |
| | | | | |
| accuracy | | | 0.88 | 7598 |
| micro avg | 0.88 | 0.88 | 0.88 | 7598 |
| macro avg | 0.87 | 0.90 | 0.88 | 7598 |
| weighted avg | 0.90 | 0.88 | 0.88 | 7598 |

**Table 1.** Classification report for the binary classification

**Multi-class Vulnerability Classification** If not only the existence of an arbitrary weakness is to be detected, but also the exact type of the respective weakness, a multi-class classification can be performed. Here, the class of non-flawed implementations remains as in the previous experiment. However, a separate class is defined for each of the flawed implementations, resulting in a total of 24 classes for the selection of weaknesses used in this training process.

As can be seen from the frequency of the classes from Table 2, there is a much greater imbalance of classes when considering each weakness individually. The basic structure of the experiment and the architectures used remain identical to the previous binary classification. The results of the training runs are shown in Figure 3.

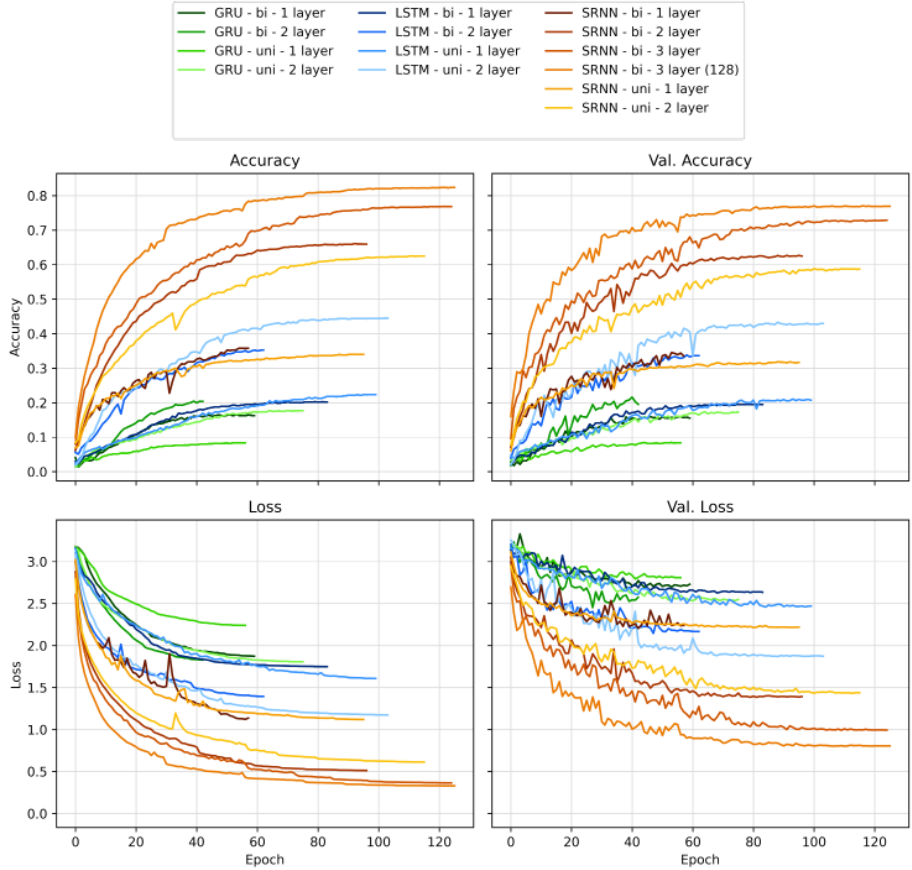

**Fig. 3.** Accuracy and loss for multi-class classification

From the training curves, clear distinctions can be made between the architectural variants in this experiment, with the SRNN implementations again producing the best results. In both the unidirectional and bidirectional implementations, it can be observed that the models with more hidden layers clearly outperform the smaller models. In a direct comparison, the models with bidirectional implementation perform better, which is why additional models were tested in this group. These are a SRNN model with three hidden layers and a unit size of 64, and a model with three hidden layers and a unit size of 128. The bidirectional SRNN with three hidden layers and the unit size of 128 gave the best results with a validation accuracy of 78% and will be used for further evaluation.

The selected model was tested again on the separate out-of-sample dataset, yielding the results of the normalised confusion matrix shown in Figure 4. Due to the large number of classes, only the numerical values in the confusion matrix that do not contain the value zero are shown to improve the readability of the figure.

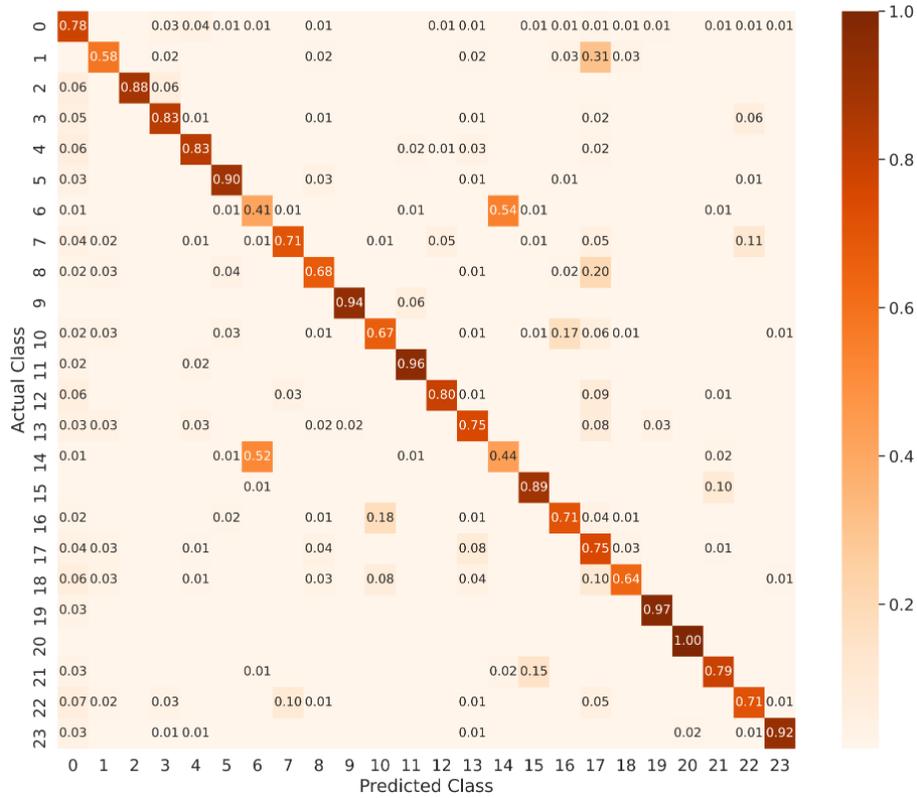

**Fig. 4.** Normalised confusion matrix for the multi-class classification

The confusion matrix shows the correct assignments of the classes, which make up the majority of all classes with minor misclassifications occurring across all classes, particularly noticeable in class 0 across all weakness classes. Other conspicuous misclassifications can be explained by the type of weakness. This applies, for example, to the Absolute Path Transversal and the Relative Path Transversal weakness (Class 6 and 14), the Integer Overflow and Integer Underflow (Class 10 and 16), and Buffer Under-read and Buffer Under-write (Class 7 and 22). These misclassifications can be explained by the similarity of these weaknesses. Further details of this classification, including the imbalance of classes, can be found in the detailed classification report in Appendix C.

## 5   Discussion

We examined whether a supervised deep learning approach using recurrent neural networks for the application of vulnerability detection based on binary executables, as introduced in [1], is a suitable means. Therefore, the existing approach was expanded in both implementation and scope. We aimed to find out if additional vulnerabilities can be detected, if the exact type of vulnerabilities can be identified and which architectural changes of a recurrent neural network are relevant for the learning process.

Extensions of this work to existing research fall into the three sub-areas of test-case selection, pre-processing, and the training process. As part of the test-case selection, we expanded the test case selection and justified this selection resulting in the use of 23 weaknesses and 50,651 samples in total. In the preprocessing step, existing methods were extended to standardise numerical values in the form of individual digits. We expect this step to be essential for the detection of memory allocation related vulnerabilities. The model training are represents our main contribution, which serves to evaluate the objectives set for this work.

Appendix C details the implementation results of our multi-classification approach. Comparing the used weaknesses CWE 134, 191, 401 and 590 of [1] with the remaining CWEs, we get a macro F1-score of 0.72 for the four CWEs and a 0.64 across all classes. The data show that the approach can be adopted to identify other types of vulnerabilities, but that on average these are more difficult to detect.

When examining the suitability of an exact identification of the weakness type, a binary classification was compared with a multi-class classification, which achieved an accuracy of 88% and 77% respectively. On the other hand, it was observed that in the case of multi-class classification, an interpretation of the error associations between related vulnerability types was shown. In addition, the learning effects and accuracy of detecting flawed code were both better with multiclass classification, so we conclude that identifying the exact type is more useful.

However, we were not able to determine from [1] whether the classifications performed consider classes in isolation and thus binary classify each of the flawed and non-flawed samples of a single CWE. We are actually concerned that such a sample selection could introduces a bias, since the isolated consideration of single vulnerabilities does not reflect the real use case of vulnerability detection, where several vulnerabilities have to be considered. For that reason our work explicitly details the the implementation of both approaches.

Regarding the influence of the architectural model decisions, in the case of binary classification, the SRNN model was found to perform best, and the number of layers and directionality had a minor influence. In the multiclass classification, the SRNN model also showed the best results, which were, however, influenced by the hyperparameters mentioned above. Both unidirectional models performed worse than the bidirectional implementation. With regard to the number of layers, it was found that a larger network produced better results. Therefore, in addition to the two-layer implementations, a three-layer implementation and another variant of this three-layer implementation with increased unit size were carried out, each of which performed better than the previous variants. From this observation, it can be concluded that the size of the network used has a significant influence compared to the previous experiment.

### 5.1   Limitations

Limitations of the approach used mainly arise from the dataset used, which consists of synthetically generated test cases of selected CWEs. It can be presumed that this data set has different properties in terms of structure and complexity than real code, which would deviate the detection rate in real use cases. Furthermore, the approach used at the function level has the disadvantage that it is not suitable for identifying weaknesses that extend across several functions.

### 5.2   Future Work

In order to overcome the limitations discussed we propose the expansion of the training dataset by using more realistic data for further enhancement. In addition a combination with taint analysis methods could be used to be able to identify vulnerabilities spreading across multiple functions. To not only identify a vulnerability, but also to find out how this vulnerability can be exploited and how the respective code location would be reached, a combination of our approach with fuzzing techniques would be possible. In the first step, possible vulnerable functions could be identified, and then fuzzing would be used to determine the most ideal way to reach them. For finding the correct execution branch, existing research has already impressively shown how the use of machine learning methods can also accelerate this process [24].

## 6   Conclusion

This paper aimed to analyse whether deep learning-based models can be used to sufficiently identify vulnerabilities in binary executables.

When reviewing existing work in the field of machine learning based vulnerability detection, it was found that the detection based on the original source code is a well researched objective that showed convincing results. In contrast, much less research has been done on static analysis of vulnerabilities using binary executables, and the work considered is limited in scope, especially with regard to the selection of vulnerability types. This motivated us to test and optimise the suitability of these existing approaches for a broader scope. Another motivation for conducting this research was the fact that almost perfect results were reported in [22] and [1], suggesting that these approaches might perform better than source code-based detection.

We have shown how the approach introduced in [1] can be extended from 4 to 23 types of identifed vulnerabilities and how additions to the preprocessing process allow for better processing of numerical values in particular. It was also pointed out how preprocessing introduces possible sources of error and how these can be removed for a clean training process. These differences in preprocessing and our extended scope meant that it was not possible to reproduce the work precisely.

We also showed that our approach can be transferred to other types of vulnerabilities and that false negatives in particular can be excluded, thus generously reducing the scope of an analysis. In the extended model selection, the findings from the work [1] regarding the result that SRNN produce the best results could be confirmed. Furthermore, the observation that two to three layer-models produces comparably good results, as reported in [15], could be confirmed. We have also shown how extending the experiment to larger unit sizes further improves the results, suggesting that there is further potential for optimisation.

Based on these findings, we can recommend the general analysis approach and state that deep learning methods are able to identify a variety of practically relevant vulnerabilities. However, in terms of practical application, it was noted that it is unclear how accurately the relationships learned from synthetic data can be recognised in real applications.

A reasonable next step for further research would therefore be to apply the investigated approach to realistic data in order to prove its added value for finding weakness with regard to real applications. For this purpose, a comparison with other methods discussed, such as the similarity-based analysis of binaries or the common methods of static analysis, would also be useful. Finally, the possibility of combining the presented approach with alternative analysis methods was discussed. We consider such a combination to be reasonable due to the possibility of precise narrowing of the scope of analysis of a binary file based on the detection methodology presented in this work.

# References

[1] J. Zheng, J. Pang, X. Zhang, X. Zhou, M. Li, and J. Wang, "Recurrent neural network based binary code vulnerability detection," in *Proceedings of the 2019 2nd International Conference on Algorithms, Computing and Artificial Intelligence*, ser. ACAI 2019, Sanya, China: Association for Computing Machinery, 2019, pp. 160–165, ISBN: 9781450372619. DOI: 10.1145/3377713.3377738. [Online]. Available: https://doi.org/10.1145/3377713.3377738.

[2] J. Li, B. Zhao, and C. Zhang, "Fuzzing: A survey," *Cybersecurity*, vol. 1, no. 1, pp. 1–13, 2018.

[3]  S. Arakelyan, S. Arasteh, C. Hauser, E. Kline, and A. Galstyan, "Bin2vec: Learning representations of binary executable programs for security tasks," *Cybersecurity*, vol. 4, no. 1, pp. 1–14, 2021.

[4]  M. Chen, J. Tworek, H. Jun, *et al.*, *Evaluating large language models trained on code*, 2021. arXiv: 2107.03374 [cs.LG].

[5]  I. U. Haq and J. Caballero, "A survey of binary code similarity," *ACM Comput. Surv.*, vol. 54, no. 3, Apr. 2021, ISSN: 0360-0300. DOI: 10.1145/3446371. [Online]. Available: https://doi.org/10.1145/3446371.

[6]  L. Nouh, A. Rahimian, D. Mouheb, M. Debbabi, and A. Hanna, "Binsign: Fingerprinting binary functions to support automated analysis of code executables," in *ICT Systems Security and Privacy Protection*, S. De Capitani di Vimercati and F. Martinelli, Eds., Cham: Springer International Publishing, 2017, pp. 341–355, ISBN: 978-3-319-58469-0.

[7]  B. Kang, T. Kim, H. Kwon, Y. Choi, and E. G. Im, "Malware classification method via binary content comparison," in *Proceedings of the 2012 ACM Research in Applied Computation Symposium*, ser. RACS '12, San Antonio, Texas: Association for Computing Machinery, 2012, pp. 316–321, ISBN: 9781450314923. DOI: 10.1145/2401603.2401672. [Online]. Available: https://doi.org/10.1145/2401603.2401672.

[8]  Z. Tai, H. Washizaki, Y. Fukazawa, Y. Fujimatsu, and J. Kanai, "Binary similarity analysis for vulnerability detection," in *2020 IEEE 44th Annual Computers, Software, and Applications Conference (COMPSAC)*, 2020, pp. 1121–1122. DOI: 10.1109/COMPSAC48688.2020.0-110.

[9]  Y. Wang, Z. Wu, Q. Wei, and Q. Wang, "Neufuzz: Efficient fuzzing with deep neural network," *IEEE Access*, vol. 7, pp. 36 340–36 352, 2019.

[10]  P. Godefroid, H. Peleg, and R. Singh, "Learn&fuzz: Machine learning for input fuzzing," in *2017 32nd IEEE/ACM International Conference on Automated Software Engineering (ASE)*, IEEE, 2017, pp. 50–59.

[11]  Z. Li, D. Zou, S. Xu, H. Jin, H. Qi, and J. Hu, "Vulpecker: An automated vulnerability detection system based on code similarity analysis," in *Proceedings of the 32nd Annual Conference on Computer Security Applications*, ser. ACSAC '16, Los Angeles, California, USA: Association for Computing Machinery, 2016, pp. 201–213, ISBN: 9781450347716. DOI: 10.1145/2991079.2991102. [Online]. Available: https://doi.org/10.1145/2991079.2991102.

[12]  J. Jang, A. Agrawal, and D. Brumley, "Redebug: Finding unpatched code clones in entire os distributions," in *2012 IEEE Symposium on Security and Privacy*, 2012, pp. 48–62. DOI: 10.1109/SP.2012.13.

[13]  Z. Liu, Q. Wei, and Y. Cao, "Vfdetect: A vulnerable code clone detection system based on vulnerability fingerprint," in *2017 IEEE 3rd Information Technology and Mechatronics Engineering Conference (ITOEC)*, 2017, pp. 548–553. DOI: 10.1109/ITOEC.2017.8122356.

[14]  S. Kim, S. Woo, H. Lee, and H. Oh, "Vuddy: A scalable approach for vulnerable code clone discovery," in *2017 IEEE Symposium on Security and Privacy (SP)*, 2017, pp. 595–614. DOI: 10.1109/SP.2017.62.

18	Andreas Schaad and Dominik Binder

[15] Z. Li, D. Zou, S. Xu, et al., "Vuldeepecker: A deep learning-based system for vulnerability detection," *Proceedings 2018 Network and Distributed System Security Symposium*, 2018. DOI: 10.14722/ndss.2018.23158. [Online]. Available: http://dx.doi.org/10.14722/ndss.2018.23158.

[16] P. E. Black et al., "Sard: Thousands of reference programs for software assurance," *J. Cyber Secur. Inf. Syst. Tools Test. Tech. Assur. Softw. Dod Softw. Assur. Community Pract*, vol. 2, no. 5, 2017.

[17] Z. Li, D. Zou, S. Xu, Z. Chen, Y. Zhu, and H. Jin, "Vuldeelocator: A deep learning-based fine-grained vulnerability detector," *IEEE Transactions on Dependable and Secure Computing*, pp. 1–1, 2021, ISSN: 2160-9209. DOI: 10.1109/tdsc.2021.3076142. [Online]. Available: http://dx.doi.org/10.1109/TDSC.2021.3076142.

[18] G. Lin, J. Zhang, W. Luo, L. Pan, and Y. Xiang, "Poster: Vulnerability discovery with function representation learning from unlabeled projects," in *Proceedings of the 2017 ACM SIGSAC Conference on Computer and Communications Security*, ser. CCS '17, Dallas, Texas, USA: Association for Computing Machinery, 2017, pp. 2539–2541, ISBN: 9781450349468. DOI: 10.1145/3133956.3138840. [Online]. Available: https://doi.org/10.1145/3133956.3138840.

[19] G. Lin, J. Zhang, W. Luo, et al., "Cross-project transfer representation learning for vulnerable function discovery," *IEEE Transactions on Industrial Informatics*, vol. 14, no. 7, pp. 3289–3297, 2018. DOI: 10.1109/TII.2018.2821768.

[20] J. Pewny, B. Garmany, R. Gawlik, C. Rossow, and T. Holz, "Cross-architecture bug search in binary executables," in *2015 IEEE Symposium on Security and Privacy*, 2015, pp. 709–724. DOI: 10.1109/SP.2015.49.

[21] S. Eschweiler, K. Yakdan, and E. Gerhards-Padilla, "Discovre: Efficient cross-architecture identification of bugs in binary code.," in *NDSS*, vol. 52, 2016, pp. 58–79.

[22] W. A. Dahl, L. Erdodi, and F. M. Zennaro, *Stack-based buffer overflow detection using recurrent neural networks*, 2020. arXiv: 2012.15116 [cs.CR].

[23] L. Cheng, Y. Zhang, Y. Zhang, et al., *Optimizing seed inputs in fuzzing with machine learning*, 2019. arXiv: 1902.02538 [cs.CR].

[24] Y. Wang, Z. Wu, Q. Wei, and Q. Wang, "Neufuzz: Efficient fuzzing with deep neural network," *IEEE Access*, vol. 7, pp. 36 340–36 352, 2019. DOI: 10.1109/ACCESS.2019.2903291.

[25] T. Boland and P. E. Black, "Juliet 1.1 c/c++ and java test suite," *Computer*, vol. 45, no. 10, pp. 88–90, 2012.

[26] V. Okun, A. Delaitre, P. E. Black, et al., "Report on the static analysis tool exposition (sate) iv," *NIST Special Publication*, vol. 500, p. 297, 2013.

[27] B. Gutstein and A. Richardson, *Juliet test suite for c/c++*, https://github.com/arichardson/juliet-test-suite-c, 2019.

[28] M. Dwarampudi and N. V. S. Reddy, *Effects of padding on lstms and cnns*, 2019. arXiv: 1903.07288 [cs.LG].

## A Number of functions per CWE

**Fig. 5.** Number of functions per CWE

## B  Selected CWEs

| Name | CWE ID | # bad | # good | # total | % of total |
|---|---|---|---|---|---|
| Heap-based Buffer Overflow | 122 | 2412 | 4127 | 6539 | 12.09% |
| Integer Overflow or Wraparound | 190 | 1441 | 3597 | 5038 | 9.32% |
| Stack-based Buffer Overflow | 121 | 1575 | 2777 | 4352 | 8.05% |
| Integer Underflow (Wrap or Wraparound) | 191 | 1125 | 2816 | 3941 | 7.29% |
| Use of Uninitialized Variable | 457 | 650 | 2320 | 2970 | 5.49% |
| Uncontrolled Format String | 134 | 820 | 2060 | 2880 | 5.33% |
| Free of Memory not on the Heap | 590 | 1141 | 1477 | 2618 | 4.84% |
| Mismatched Memory Management Routines | 762 | 611 | 1940 | 2551 | 4.72% |
| Improper Neutralization of Special Elements used OS Command Injection | 78 | 960 | 1260 | 2220 | 4.11% |
| Relative Path Traversal | 23 | 930 | 1230 | 2160 | 3.99% |
| Buffer Underwrite ('Buffer Underflow') | 124 | 810 | 1333 | 2143 | 3.96% |
| Absolute Path Traversal | 36 | 907 | 1205 | 2112 | 3.91% |
| Unexpected Sign Extension | 194 | 907 | 983 | 1890 | 3.50% |
| Signed to Unsigned Conversion Error | 195 | 896 | 988 | 1884 | 3.48% |
| Buffer Under-read | 127 | 764 | 1110 | 1874 | 3.47% |
| Improper Release of Memory Before Removing Last Reference (Memory Leak) | 401 | 265 | 1455 | 1720 | 3.18% |
| Uncontrolled Resource Consumption ('Resource Exhaustion') | 400 | 543 | 1154 | 1697 | 3.14% |
| Divide By Zero | 369 | 535 | 1107 | 1642 | 3.04% |
| Buffer Over-read | 126 | 602 | 910 | 1512 | 2.80% |
| Integer Overflow to Buffer Overflow | 680 | 450 | 474 | 924 | 1.71% |
| Double Free | 415 | 181 | 393 | 574 | 1.06% |
| Numeric Truncation Error | 197 | 447 | 54 | 501 | 0.93% |
| Unchecked Return Value to NULL Pointer Dereference | 690 | 94 | 234 | 328 | 0.61% |

**Table 2.** Selected samples per CWE

## C   Multi-class Classification Report

| Class ID | CWE ID | Precision | Recall | F1-score | Support |
|---|---|---|---|---|---|
| 0 | - | 0.98 | 0.78 | 0.87 | 5004 |
| 1 | 197 | 0.55 | 0.58 | 0.56 | 62 |
| 2 | 401 | 0.71 | 0.88 | 0.78 | 33 |
| 3 | 121 | 0.58 | 0.83 | 0.68 | 237 |
| 4 | 122 | 0.55 | 0.83 | 0.66 | 333 |
| 5 | 194 | 0.64 | 0.90 | 0.75 | 88 |
| 6 | 23 | 0.35 | 0.41 | 0.38 | 127 |
| 7 | 127 | 0.67 | 0.71 | 0.69 | 126 |
| 8 | 195 | 0.58 | 0.68 | 0.62 | 103 |
| 9 | 415 | 0.76 | 0.94 | 0.84 | 17 |
| 10 | 190 | 0.67 | 0.67 | 0.67 | 187 |
| 11 | 762 | 0.72 | 0.96 | 0.82 | 93 |
| 12 | 126 | 0.49 | 0.80 | 0.61 | 89 |
| 13 | 680 | 0.40 | 0.75 | 0.52 | 61 |
| 14 | 36 | 0.37 | 0.44 | 0.40 | 110 |
| 15 | 78 | 0.66 | 0.89 | 0.76 | 130 |
| 16 | 191 | 0.59 | 0.71 | 0.64 | 162 |
| 17 | 400 | 0.24 | 0.75 | 0.36 | 72 |
| 18 | 369 | 0.59 | 0.64 | 0.61 | 77 |
| 19 | 457 | 0.53 | 0.97 | 0.68 | 79 |
| 20 | 690 | 0.36 | 1.00 | 0.53 | 13 |
| 21 | 134 | 0.48 | 0.79 | 0.60 | 94 |
| 22 | 124 | 0.48 | 0.71 | 0.57 | 119 |
| 23 | 590 | 0.80 | 0.92 | 0.86 | 182 |
| **accuracy** | | | | 0.77 | 7598 |
| **macro avg** | 0.57 | | 0.77 | 0.64 | 7598 |
| **weighted avg** | 0.84 | | 0.77 | 0.79 | 7598 |

**Table 3.** Classification report for the multi-class classification